\documentclass[essd,preprint]{copernicus}

\usepackage{svg}
\usepackage{makecell}

\usepackage{xurl}

\begin{document}

\title{Settlement percolation: global maps of Critical Distances}

\Author[1]{Martin}{Schorcht}
\Author[1][m.behnisch@ioer.de]{Martin}{Behnisch}
\Author[2,3]{Larissa T.}{Beumer}
\Author[1,4]{Anna-Katharina}{Brenner}
\Author[1]{Renan L.}{Fagundes}
\Author[1]{Tobias}{Kr\"uger}
\Author[3,5]{Thomas}{M\"uller}
\Author[3,6]{Wenjing}{Xu}
\Author[1,7][ca-dr@rybski.de]{Diego}{Rybski}


\affil[1]{Leibniz Institute of Ecological Urban and Regional Development (IOER), Dresden, Germany}
\affil[2]{The University Centre in Svalbard (UNIS), Longyearbyen, Norway}
\affil[3]{Senckenberg Biodiversity and Climate Research Centre (SBiK-F), Frankfurt am Main, Germany}
\affil[4]{Institute of Social Ecology, BOKU University, Vienna, Austria}
\affil[5]{Department of Biological Sciences, Goethe University Frankfurt, Frankfurt am Main, Germany}
\affil[6]{University of Massachusetts Amherst, Amherst, MA, USA}
\affil[7]{Complexity Science Hub Vienna, Metternichgasse 8, A-1030 Vienna, Austria}

\nolinenumbers

\runningtitle{Settlement percolation: global maps of Critical Distances}

\runningauthor{M.\ Schorcht et al.}

\received{}
\pubdiscuss{}
\revised{}
\accepted{}
\published{}

\firstpage{1}

\maketitle

\begin{abstract}
A substantial share of the Earth’s land surface is managed by humans, with cities representing the most extreme form of anthropogenic land use. There are zillion ways in which settlements can be arranged across a given area, and their specific spatial configuration has important consequences for both urban systems and the natural environment. Here, we introduce a novel approach to characterizing settlement configuration by systematically quantifying it in terms of a transition resembling percolation -- that is, by identifying the critical distance at which isolated settlements merge into a giant, overarching settlement cluster. We estimate this critical distance across multiple spatial scales and units, including national and subnational levels, non-overlapping tiles, and moving windows, covering the entire globe. The critical distance provides an independent measure of settlement connectivity and thus adds value to spatial analyses of settlement structure and its social, economic, and ecological impacts. Accordingly, our Global Settlement Percolation (GSP) dataset is relevant to a wide range of research communities, including those studying urban morphology, land-use patterns, and landscape ecology.
\end{abstract}

\introduction

The configuration of human settlements is central to equitable access to essential social services, public transport, green spaces, and housing. At the same time, human settlements also contribute to the fragmentation of natural habitats, affecting  ecological processes such as water flows and wildlife movement and hence ecological connectivity that is fundamental for biodiversity. Therefore, understanding and managing sustainable human development or ecosystem dynamics and biodiversity rely on the ability to quantitatively characterize settlement configuration.

The physical properties of the settlement configuration are fundamental characteristics of urban systems, and many studies seeking to understand urban growth, functionality, and impacts on ecosystems aim to describe these properties \citep{georgNewSpatialDimensions2016,cushmanParsimonyLandscapeMetrics2008}. Settlement configurations are the physical result of numerous interacting decisions and processes, and in their physical form, settlements are interlinked and exhibit varying degrees of coalescence \citep{georgNewSpatialDimensions2016}. These processes are inherently non-linear and give rise to emergent spatial patterns, settlements often exhibit self-similar patterns of agglomeration and clustering across spatial scales \citep{BattyM2008}, making it possible to describe them using concepts and theoretical frameworks from statistical physics and complex systems science.  

Existing local and regional studies on settlement configuration have been inspired by concepts from urban geography such as Central Place Theory \citep{centrealPlaceEnglish2019}, urban sprawl \citep{chettryCriticalReviewUrban2023,behnisch_sprawl_2022}. 
High-resolution global patterns of settlement agglomeration and dispersion have been examined using, for example, gravitational approaches \citep{StranoSNEM2021,RybskiRK2013}. 
Approaches addressing human connectivity often focus on transport systems, internet-based datasets e.g. \citep{tianFlowSpaceReveals2023}, or, more recently, night-time lights \citep{lvAdvancingUrbanConnectivity2025}. 
However, no globally consistent dataset currently exists that quantifies the connectivity of settlement systems from their spatial configuration across multiple spatial scales. 

\begin{figure*}[t]
\centering
\includegraphics[width=0.65\textwidth]{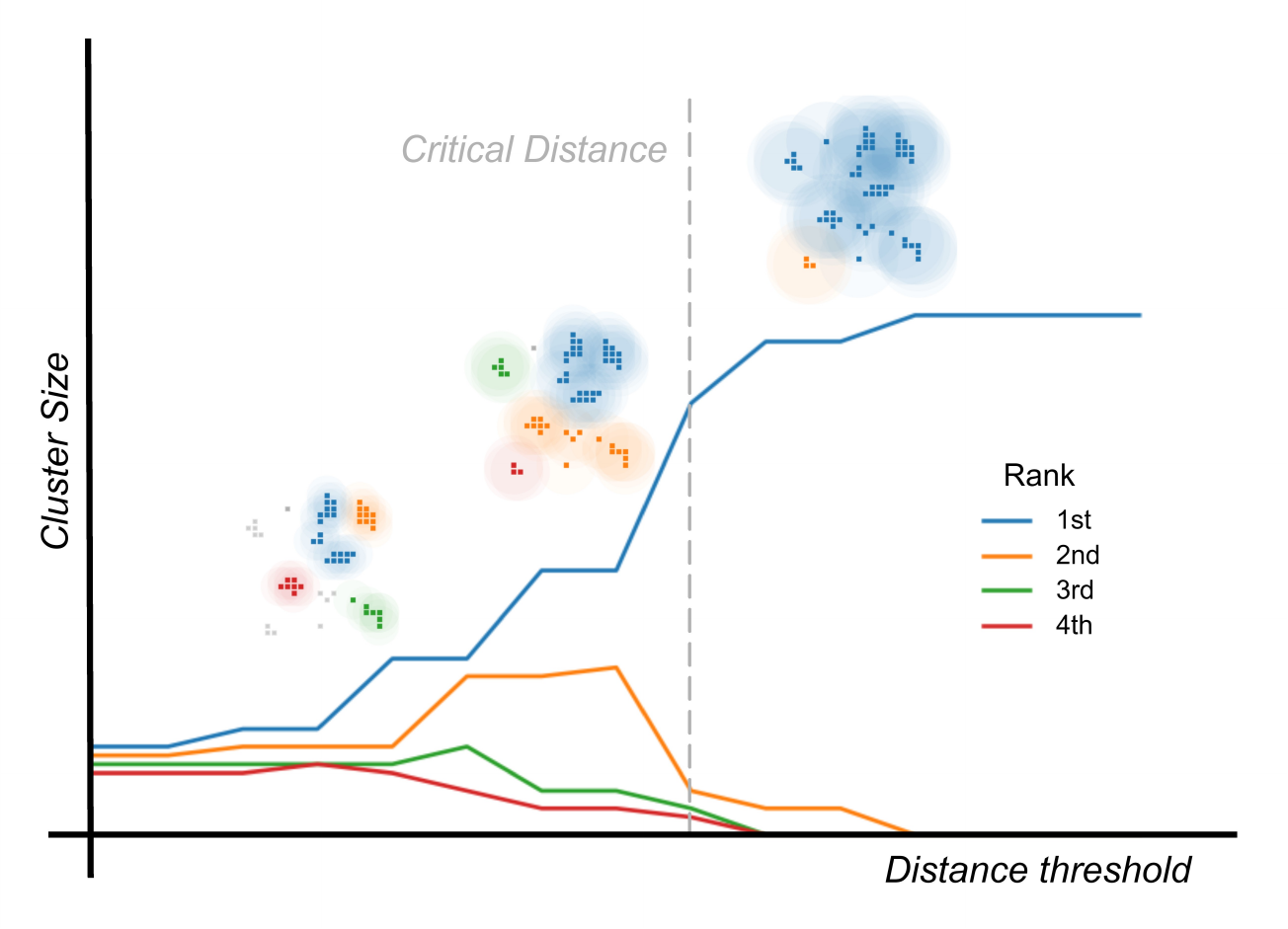}
\caption{Illustration of clustering and percolation-like transition.
Any two pixels within a predefined distance threshold are assigned to the same cluster.
Choosing a small threshold (left), one obtains many small clusters.
Increasing the distance threshold, clusters merge, and for a large threshold (right) most pixels belong to the largest cluster.
The colored lines represent the sizes of the first four largest clusters.
One can see that the largest one grows and at the critical distance (dashed line) abruptly dominates and becomes the giant cluster.
At the same time the second largest suddenly shrinks (actually, the second largest merges with the largest and another cluster becomes second largest).
Figures~\ref{fig:crit_dist_example} and~\ref{fig:PERr106} illustrate the transition with real-world data.
}
\label{fig:conceptual}
\end{figure*}

Defining spatial clusters by a distance threshold can be used to analyze \emph{settlement percolation}.
By assigning any two points or cells to the same cluster if their distance is less than a threshold $l$, one obtains many small clusters for small thresholds (Fig.~\ref{fig:conceptual}, left) and a giant overarching cluster for large thresholds (Fig.~\ref{fig:conceptual}, right).
In between these two extremes, a transition occurs from a fragmented state to a connected state.
This phase transition shares characteristics with phenomena that are being studied under the umbrella of percolation theory \citep[e.g.]{BundeHFractalsDisorderedSystems1991-2,StaufferA1994}.
The transition is reflected in the cluster sizes (Fig.~\ref{fig:conceptual}), which can be utilized to identify the Critical Distance.
The Critical Distance is the clustering threshold at which a dominant connected settlement cluster emerges (commonly referred to as a \emph{giant cluster} \citep[e.g.]{FagundesLRR2025}). 
In qualitative terms, the Critical Distances can be understood as the bottleneck between settlements, and larger values indicate more dispersed settlement patches, whereas smaller Critical Distances indicate more closely spaced, clustered settlement patches.
In other words, the Critical Distance is a measure of configuration that is complementary to composition.
The same number of urban pixels, can have different Critical Distances depending on how these pixels are arranged across the unit.

For cities, settlements, and urban systems, variants of settlement percolation have been analyzed in various contexts \citep{FagundesLRR2025}.
\cite{ArcauteHFYJB2015,ArcauteMHMVRMB2016} analyze the percolation of Britain's street network and its intersections to obtain a hierarchical structure between regional fractures and city morphology.
\cite{FluschnikKRZRKR2016} apply a similar methodology to urban land cover and estimate the Critical Distance for a set of countries. 
E.g.\ Austria's settlements are in a fragmented phase for distances below 15\,km and in a connected one above this Critical Distance.
\cite{HuynhMLMC2018} analyze the transport points (bus stops) and consider the characteristic distance and the respective cluster area.
Additionally, employing the respective standard deviations, the authors define a phase space separating clustered and dispersed patterns \citep{HuynhMLMC2018}.
\cite{behnisch_settlement_2019} apply spatial clustering to the German building stock and estimate a Critical Distance of only 830\,m.
These recent works go beyond the neutral models proposed by \cite{GardnerMTN1987} -- but are limited to case studies and a global analysis is lacking.
For an overview we refer to \cite{FagundesLRR2025}.

Building on the settlement percolation approach, we fill this gap and systematically analyze high-resolution urban land cover of the whole globe and provide the Global Settlement Percolation (GSP) dataset consisting of Critical Distances for different spatial units and scales.
We use the World Settlement Footprint (WSF2019) as input data, a globally consistent, high-resolution settlement dataset derived from Sentinel-1 (S1) and Sentinel-2 (S2) that maps buildings and roofed infrastructure at 10\,m resolution across the globe.
We perform spatial clustering of all settlement patches at this 10\,m resolution and determine the Critical Distances at which settlement pixels transition from a disconnected to a connected state for different spatial units -- including countries and sub-national regions -- and for various grid cell sizes (0.5$^\circ$, 1$^\circ$, 2.5$^\circ$, 5$^\circ$). For 1\,km grids, we apply a moving window approach with varying radii (1, 2.5, 5, 7.5, 10, and 15\,km) to capture settlement percolation at a fine scale. 

As a novel measure for quantifying settlement configuration, this dataset can be used in downstream research to support large-scale comparative analysis, for example to describe different patterns of urban morphology and to characterize varying human mobility needs for travel between settlements.
The connectivity of settlements, at the same time, informs about the fragmentation of the non-urban landscape.
Accordingly, our data could also be used to characterize the non-built up areas in terms of landscape ecology. 
Moreover, the Critical Distance as provided by our work can be informative about how close the urban system is to merging into a giant object, i.e.\ how close are neighboring cities to merge into a megapolis.
In sum, the GSP dataset has broad applicability for various disciplines, including spatial planning, urban studies, landscape architecture, (landscape) ecology, and conservation science.

\section{Methods}
\subsection*{Input-Data}

\subsubsection*{World Settlement Footprint (WSF)}
We use the \emph{World Settlement Footprint} (\emph{WSF}) dataset from 2019 \citep{marconcini_understanding_2021} as input data for the global built-up areas, which in the following we consider as \emph{settlement areas}.
The \emph{WSF} is a one-band binary raster dataset with approximately 4 million horizontal by 1.5 million vertical pixels, with a pixel size of 10\,m at the equator (Geographic Coordinate System is WGS84).  
The uncompressed size is 5.6 \,TB, with deflate compression it is 19.3 \,GB.  
The \emph{WSF} was derived and provided by the \emph{German Aerospace Center} \citep{wsf2019DownloadPage} from \emph{Sentinel-1} and \emph{Sentinel-2} satellite images. 
The \emph{WSF} does not include the road network. 
The accuracy is very high at up to 89\,\%, at least for the older dataset from 2015 \citep{Marconcini2020}.
Thus, this dataset is very suitable for deriving spatial clusters of settlement structures, in particular at a global scale.

\subsubsection*{Additional data}
To determine the proportion of settlement areas in relation to terrestrial areas, we use the \emph{ESRI World Water Bodies} dataset \citep{esri_world_2023}. 
It includes the world's rivers, lakes, dry salt pans, seas, and oceans. 
However, smaller bodies of water and rivers are not distinguished. 

Furthermore, we use administrative boundaries of countries, states and provinces from the \emph{Natural Earth} dataset \citep{naturalearthAdministrativeAreas2022} to calculate the results for territorial geometries.

\subsection*{Analysis}\phantomsection\label{subsec:analysis} 
As a first step, we conduct spatial clustering of all settlement patches worldwide.
We then determine the Critical Distance of settlement clusters for each resolution and unit.
The Critical Distance denotes the clustering threshold at which settlement patterns transition from a fragmented to a connected configuration (or vice versa), marked by the emergence of a dominant giant cluster, resembling a classical percolation transition \citep[e.g.,][]{FagundesLRR2025,behnisch_settlement_2019,FluschnikKRZRKR2016}.
We apply clustering according to the geodetic distance \citep{RozenfeldRABSM2008,RozenfeldRGM2011}, taking into account the curvature.

The section is structured into data preprocessing, identification of the clusters based on various distances, and finally the calculation of the Critical Distance of specific spatial units. Furthermore, detailed information regarding technical implementation is explained in a separate sub-section.      
An overview of the entire workflow is shown in Fig.~\ref{fig:workflow}.

\subsubsection*{Data preprocessing}
The global \emph{WSF 2019} dataset, obtained from \citet{wsf2019DownloadPage}, comprises 5,138 GeoTIFF raster tiles, each containing 22,488 columns and 22,487 rows. 
Because adjacent raster tiles overlap along their boundaries, these overlapping regions were removed prior to further processing. 
To achieve this, all input tiles were first merged into a seamless global raster, thereby eliminating duplicate edge pixels. 
The resulting global raster was subsequently partitioned into tiles of 50,000 × 50,000 pixels, a size selected to balance computational efficiency, memory requirements, and parallel processing performance.

\subsubsection*{Clustering with distance threshold $l$}
\begin{figure*}[t]
\centering
\includegraphics[width=0.8\textwidth]{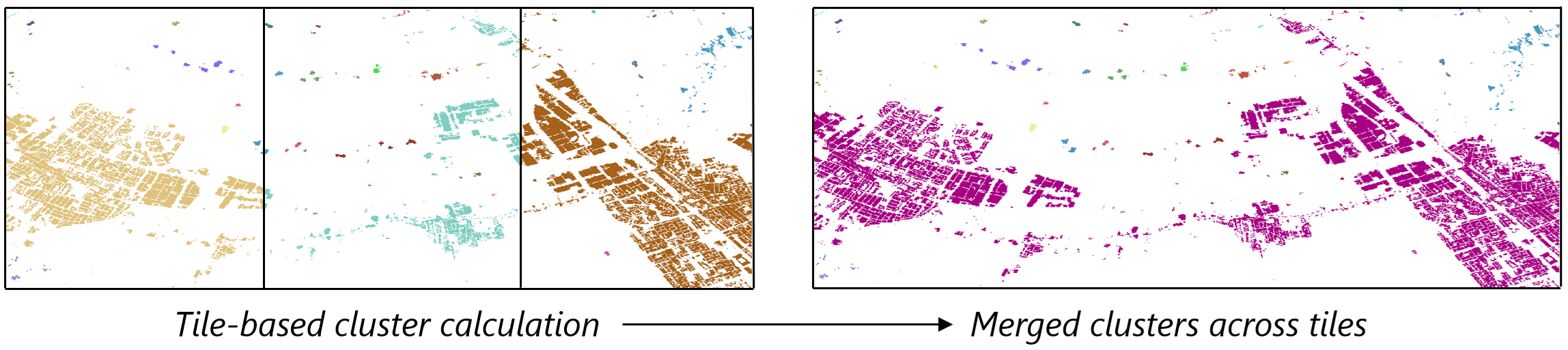}
\caption{Illustration of how clusters in tiles can be connected across tiles.
For performance reasons, spatial clustering via a distance threshold had to be done in separate tiles.
Thus, each of the three boxes on the left contains independent clusters.
However, clusters can extend beyond the limits of these tiles, so that in a next step they need to be consolidated.
Then, on the right, the purple cluster extends across the entire area (which previously consisted of three tiles).
}
\label{fig:tilebased_clustering}
\end{figure*}

We apply spatial clustering, i.e.\ all settlement cells that are within the predefined distance $l$ from each other are assigned to the same cluster \citep{RozenfeldRABSM2008,RozenfeldRGM2011}. 
We use the \texttt{PostgreSQL} extension \texttt{PostGIS} for clustering.
The amount of data to be processed was so large that \texttt{PostgreSQL} could not process the worldwide clustering at once. 
Accordingly, we divide the datasets into tiles during preprocessing (see above).
In a first step, we carry out a cluster process for each of these tiles individually.
In a further step, we consolidate the resulting clusters of the individual tiles to obtain clusters that can span multiple tiles (see Fig.~\ref{fig:tilebased_clustering}). 
This two-stage tile-based clustering algorithm has the advantage that significantly fewer operations are required than when the entire dataset is clustered in a single pass.
Furthermore, the calculation can be parallelized in parts, saving computing time. 
More detailed information can be found in the \hyperref[subsec:implementation]{Technical implementation} section.

In total, we perform the global clustering process for 60 different clustering thresholds $l$, with iteration steps increasing for larger distances.
We sample discrete distance thresholds from the (in principle) continuous distance variable.
The list of the various clustering thresholds is shown in Tab.~\ref{tab:clusterSteps}. 
On average, the cluster calculation of one clustering threshold took around 6~hours, whereby the calculation time increased with increasing clustering threshold. 
For example, the runtime was 4~hours for $l$ = 100\,m and 28~hours for $l$ = 30,000\,m. 

\begin{table*}[t]
\centering
  \begin{tabular}{|l|l|l|}
    \hline
    \thead{Range [m]} & \thead{Step size [m]} & \thead{Clustering thresholds [m]} \\
    \hline
    50 - 1,000 & 50 & 50; 100; \dots{}950; 1,000 \\
    \hline
    1,000 - 5,000 & 250 & 1,000; 1,250; \dots{}4,750; 5,000 \\
    \hline
    5,000 - 10,000 & 500 & 5,000; 5,500; \dots{}9,500; 10,000 \\
    \hline
    10,000 - 20,000 & 1,000 & 10,000; 11,000;\dots{}19,000; 20,000 \\
    \hline
    20,000 - 30,000 & 5,000 & 20,000; 25,000; 30,000 \\
    \hline
  \end{tabular}
\caption{\label{tab:clusterSteps} Sampling of distance thresholds used for spatial clustering.
In order to identify the Critical Distance, i.e., the distance at which a giant cluster emerges, the distance threshold which defines the spatial clusters needs to be varied.
This table lists which distance thresholds we use.
For different ranges (left), different step sizes (middle) have been chosen, leading to sets of thresholds (right).
As can be seen, the increments have been chosen in a way where for larger thresholds also the increments become larger.
The overall number of thresholds is limited by computational time.
}
\end{table*}

\subsubsection*{Identifying Critical Distances $l_\text{c}$}
\begin{figure*}[t]
\centering
\includegraphics[width=0.9\textwidth]{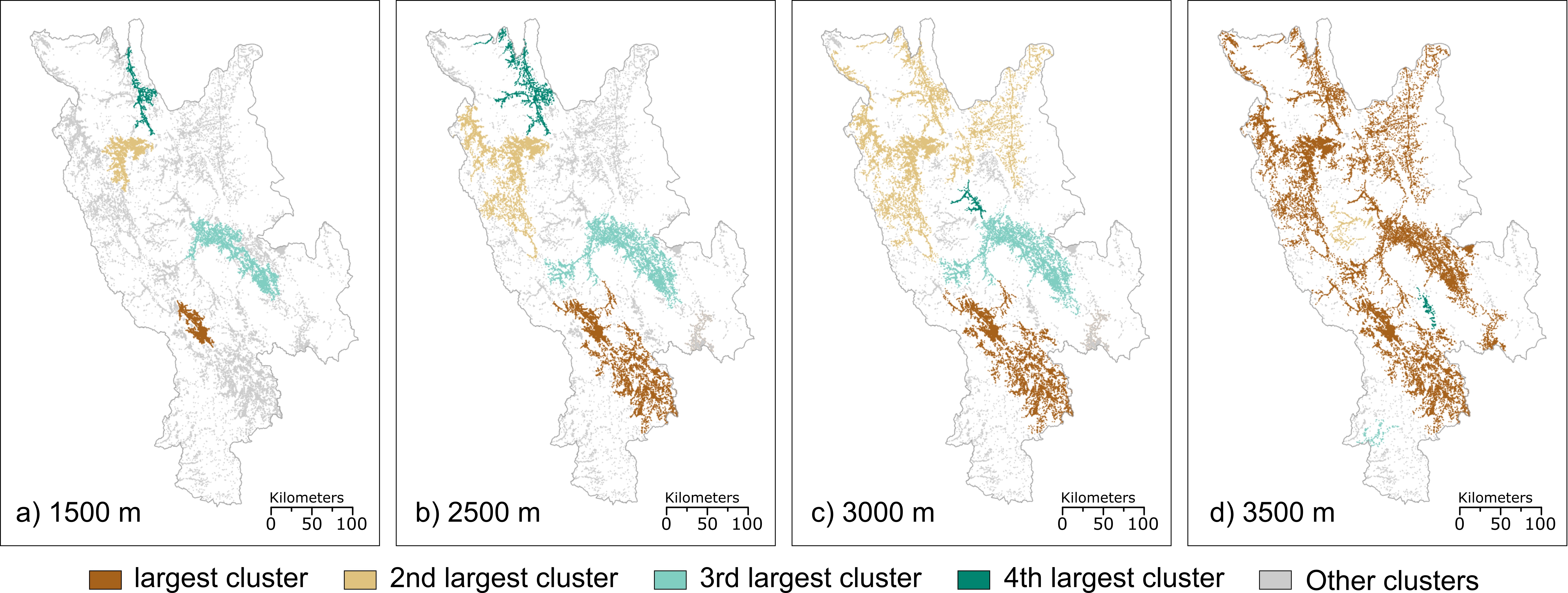}
\caption{Example of the percolation-like transition of settlements.
The maps show the settlement areas in central Peru (Huancavelica, Huanuco, Junin, and Pasco regions) where the largest four clusters are colored.
The four panels exhibit the clusters for the distance thresholds (a) $l=$1,500\,m; (b) $l=$2,500\,m; (c) $l=$3,000\,m; (d) $l=$3,500\,m.
With increasing thresholds the clusters grow, panels~(a)-(c), but beyond $l=$3,500\,m the largest cluster dominates, panel~(d), while remaining clusters are relatively small.
The respective cluster sizes are plotted in Fig.~\ref{fig:PERr106}.
}
\label{fig:crit_dist_example}
\end{figure*}

We measure the Critical Distance $l_\text{c}$ by identifying the clustering threshold at which the giant cluster is formed and smaller clusters reach their maximum size. 
An example of cluster growth as the clustering threshold increases is shown in Fig.~\ref{fig:crit_dist_example}, which represents the central part of Peru (including the regions Huancavelica, Huanuco, Junin, and Pasco).
The four largest clusters (brown, beige, turquoise, and green) are shown for each clustering threshold. 
It can be seen that, as the clustering threshold increases up to 3,000\,m, the clusters continue to grow separately. 
However, at a clustering threshold of 3,500\,m, the transition occurrs, resulting in a single dominant giant cluster.
Although there are still some settlement areas that are not part of the largest cluster, they are small and the giant cluster dominates.

\begin{figure}[t]
\centering
\includegraphics[width=0.9\columnwidth]{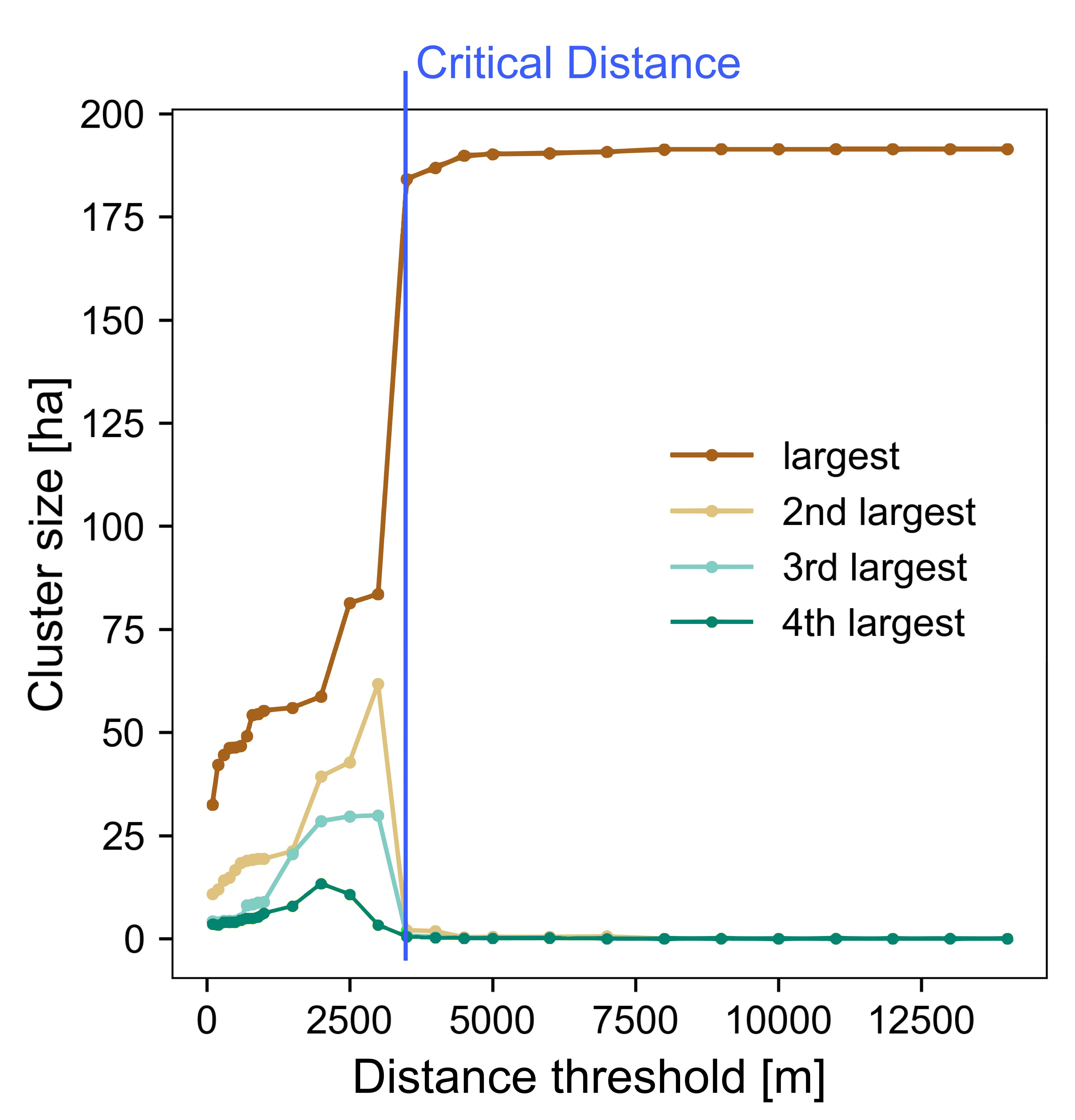}
\caption{Example of Critical Distance identification.
For central Peru (Huancavelica, Huanuco, Junin, and Pasco regions) the sizes of the largest four clusters, as shown in Fig.~\ref{fig:crit_dist_example}, are plotted against the distance threshold.
It can be seen that with increasing distance threshold the clusters grow.
This growth happens because (i) more points are assigned to the clusters and (ii) the merging of clusters.
At the Critical Distance (blue vertical line) there is a phase transition resembling percolation where the largest cluster dominates and the smaller clusters become negligible.
The underlying cluster size was determined based on the calculated geodetic area rather than the number of pixels.
}
\label{fig:PERr106}
\end{figure}

In order to determine a specific clustering threshold at which the largest clusters merge to form a giant cluster, we carry out a separate analysis for each spatial unit examined. 
Therefore, we calculate the average cluster area of the second to fourth largest clusters for each threshold distance.
The clustering threshold at which this average area decreases the most is considered the Critical Distance (to be specific, we take the smallest distance at which the giant cluster is formed, see Fig.~\ref{fig:PERr106}).
This is because if the area of the second to fourth largest clusters decreases abruptly, they have merged to form a giant cluster.
Figure~\ref{fig:PERr106} illustrates, for the same example as in Fig.~\ref{fig:crit_dist_example}, the development of the cluster areas for the various clustering thresholds.
Between 3,000\,m and 3,500\,m the largest cluster (brown) grows considerably, while the other clusters decrease in size. 

In rare cases, it can happen that the objects in an investigation area already form a single cluster at the lowest clustering threshold of 50\,m (which means the Critical Distance is 50\,m or lower). 
This occurs, for example, if the area is very densely built or if there are very few objects lying next to each other.
The larger the investigation areas are, the less likely it is that there is only one cluster at the lowest clustering threshold.
On the contrary, it can also happen that no giant cluster has formed up to the largest clustering threshold of 30\,km and therefore no transition takes place, as the individual clusters are further apart than the largest threshold.
This is more likely to occur in larger and less densely populated areas.

The Critical Distance employs the clustering threshold, which is most commonly used in the urban context \citep{FagundesLRR2025}.
To be more specific, the Critical Distance measures the connectivity of settlements, but these settlements do not necessarily extend over the entire spatial unit.
The criterion is the emergence of a giant settlement cluster.

\subsubsection*{Considered spatial units and scale}
Based on the spatial clustering, the Critical Distance can be calculated for pre-defined spatial units.
Depending on the particular usage, various spatial units may be of interest, so we calculated the Critical Distance for administrative areas, (non-overlapping) grid cells, and raster-based moving windows (see Tab.~\ref{tab:files}). 
In addition, for every spatial unit we also have different spatial scales, representing different granularities.

For administrative units, we provide the Critical Distances for 258~countries at the national level and 4,596~administrative divisions at the sub-national level, such as federal states or provinces. 
If a country or sub-nation consists of several polygons (e.g.\ in the case of islands), we provide a separate Critical Distance for each polygon, 
as otherwise the distance between the islands would have an influence on the Critical Distance.

Furthermore, we provide estimates in global grid cells (tiles) as spatial units with resolution of 0.5, 1, 2.5 and 5 degrees (see Tab.~\ref{tab:files}), which are frequently used by other studies and allow for combination with other datasets.
Grid cells whose land mass is divided by larger bodies of water (such as the English Channel) are also split into separate polygons, so that we provide an own value for each land mass. 

\begin{figure}[t]
\centering
\includegraphics[width=0.75\columnwidth]{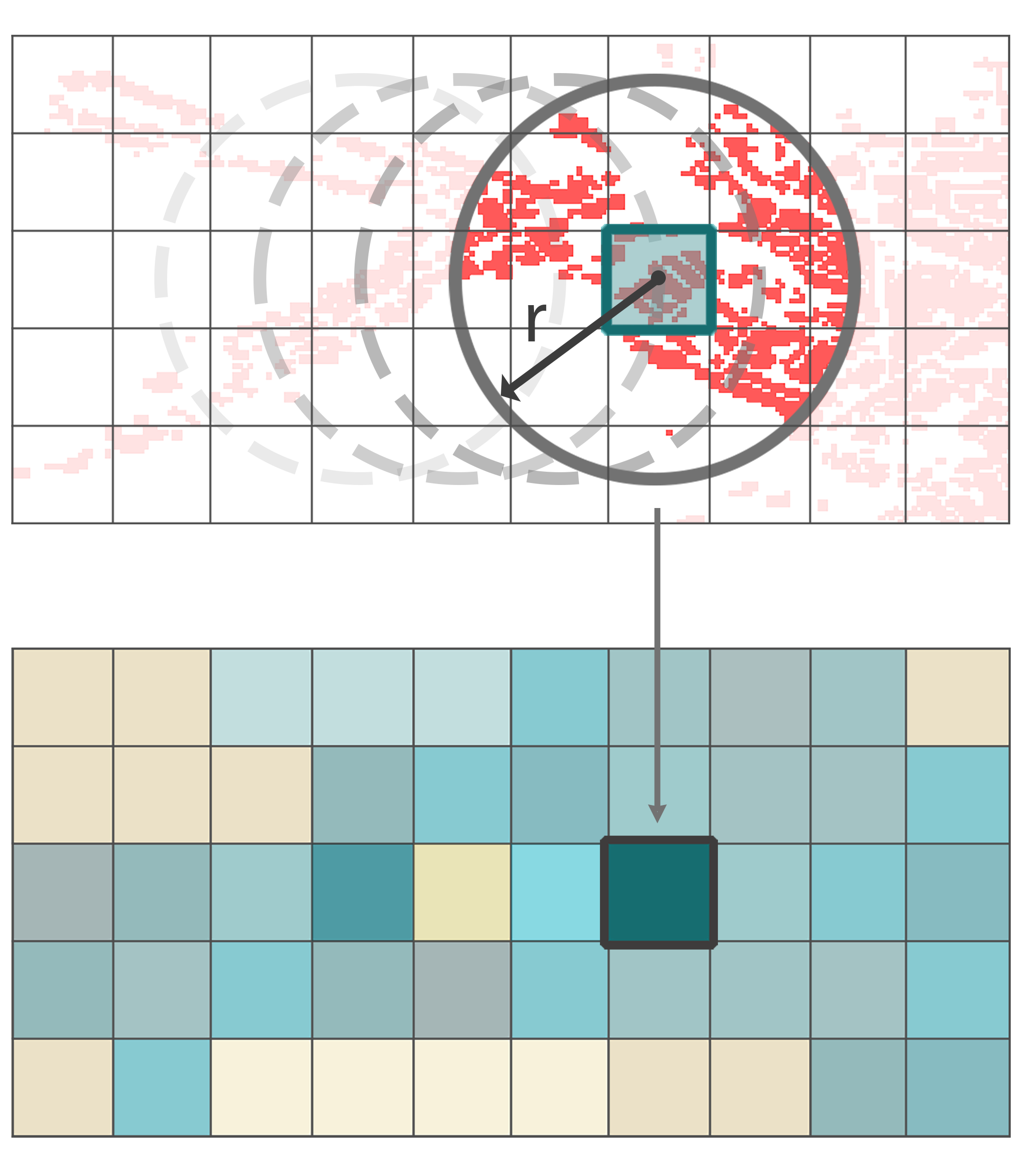}
\caption{Illustration of the moving window analysis.
The Critical Distance of the settlement areas (red) is determined within a circle of a predefined radius (top).
The resulting value is assigned to the raster cell in the center of the circle.
The circle is then shifted to the next cell and the procedure is repeated.
If the respective circles are non-empty, every cell obtains an estimate (bottom).
We use a sampling resolution of 1\,km.
In contrast to the grid cell analysis, the circles overlap (see Tab.~\ref{tab:files}).
}
\label{fig:moving_window_example}
\end{figure}

We also implement a moving window version, which adds a higher level of detail to the grid cells described in the previous paragraph.
To do so, we create a raster with a resolution of around 0.013~degrees (which approximately corresponds to a resolution of 1\,km near the equator).
A circle of a given radius around the center of each pixel represents the area of interest (spatial unit).
The Critical Distance is then calculated for each circle and written as a value in the pixel. 
We repeat the procedure for all pixels while the respective circles usually overlap.
We define six radii between 1\,km and 15\,km as spatial scale for the moving window analysis (see Tab.~\ref{tab:files}), so that six rasters are generated (one resulting raster for each radius). 
The circles are formed using geodetic buffers and are therefore also almost distortion-free. 
These rasters carry local information on the criticality of settlements within the various radii.
This procedure is shown schematically in Fig.~\ref{fig:moving_window_example}. 
The Critical Distance is calculated for this selected circular area and assigned to the respective pixel. 
If there are no settlement areas in a circle, no Critical Distance can be calculated, which is why a \texttt{nodata} value is assigned in this case. 
With larger radii, the probability that settlement areas are present within a circular area increases, which reduces the number of \texttt{nodata} pixels.

\subsubsection*{Technical implementation}\phantomsection\label{subsec:implementation}

The data were stored and processed using \texttt{PostgreSQL~15} with the \texttt{PostGIS~3.4.0} extension, while workflow execution and process orchestration were implemented in \texttt{Python~3.9}.
Data import and export operations were performed using \texttt{GDAL~3.2.2}.
To ensure the reproducibility and portability of the proposed methodology, all software components and custom scripts were integrated into a \texttt{Docker\,Compose} environment.
In addition, \texttt{PGAdmin} was included to facilitate database management and data exploration.
This containerized environment can be readily applied to compute spatial clusters and the Critical Distance for very large raster and vector datasets beyond those considered in this study (see: \url{https://github.com/schorcht-ioer/critical_mapper}).

Computationally intensive tasks were parallelized using Python's \texttt{multiprocessing pool}, with parallel execution based on tiled partitions of the input data to improve processing efficiency and scalability.

During data import, the WSF 2019 raster dataset was converted into vector geometries using the \texttt{PostGIS} function \texttt{ST\_DumpAsPolygons}.
In this process, directly adjacent raster pixels connected through a four-neighborhood were merged into single polygons, reducing approximately 8.8 billion settlement pixels to approximately 524 million polygon features.
The resulting vector dataset occupied approximately 300 GB of uncompressed disk space.
This vectorization step was required because the spatial clustering algorithms implemented in \texttt{PostGIS} operate exclusively on vector geometries.

Although vectorization introduces an additional preprocessing step, representing the data as vector geometries enables the use of spatial indexes, resulting in substantially improved computational performance.

The clustering functions provided by \texttt{PostGIS} operate exclusively on projected (planar) geometries.
To account for projection-induced geometric distortions when processing the global dataset, a geodetic buffer was generated for each settlement polygon using \texttt{ST\_Buffer(geography)}.
Internally, each geometry is transformed into an appropriate local projection prior to buffering, ensuring that the specified buffer distance closely approximates the corresponding geodetic distance while retaining the computational efficiency of planar operations.
This approach provides a favorable balance between geometric accuracy and processing performance.

Following the buffering step, polygons were assigned to clusters using the \texttt{ST\_ClusterIntersectingWin(geometry)} function, which groups all intersecting buffered geometries into connected components.
Clustering was initially performed independently for each spatial tile in parallel.
In a subsequent merging step, clusters located on adjacent tile boundaries were combined whenever their buffered geometries intersected, as determined by \texttt{ST\_Intersects(geometry)}.
This merging step, which cannot be parallelized because it requires comparisons across tile boundaries, produces seamless clusters spanning multiple tiles.

The resulting two-stage clustering workflow substantially reduces memory requirements while enabling parallel execution of the computationally most demanding processing steps.

After tile-spanning clusters have been generated for each of the 60 clustering distance thresholds, the Critical Distance is determined. For this purpose, the geodetic area of all polygons belonging to each cluster is computed using the \texttt{ST\_Area(geography)} function. These area measurements provide the basis for the subsequent determination of the Critical Distance.

\section{Results}
As detailed in the previous section (\hyperref[subsec:analysis]{Analysis}), we provide Critical Distance estimates for a range of spatial units and scales.
The taxonomy is listed in Tab.~\ref{tab:files}.
The \emph{Global Settlement Percolation} consists of three components, namely the \emph{administrative units product}, the \emph{grid cells product}, and the \emph{moving window product}.
Each product consists of various files corresponding to different spatial scales, as detailed in Tab.~\ref{tab:files}. 
All geodata are stored in the geographic coordinate system \texttt{WGS 84} (\texttt{EPSG: 4326}).
For the products in vector format (\emph{Administrative units} and \emph{Grid cells}) the files are in \texttt{Geopackage} format (\texttt{gpkg} extension), which contain the individual layers.
For example the \emph{administrative units product} consists of a \emph{national layer} and a \emph{sub-national layer}.
The respective attributes are described in Tab.~\ref{tab:attributes}.
In addition to the Critical Distance and attributes, the vector data has a \texttt{status} attribute, which describes whether percolation has taken place (\texttt{possible values: percolated, not percolated, nodata}). 
The meaning of \texttt{not percolated} is that no giant cluster was formed, as either the objects were already grouped into one cluster at the smallest clustering threshold, or the largest clustering threshold was not large enough to group the objects into a giant cluster.
The \texttt{status} value is set to \texttt{nodata} if there is no settlement area in a spatial unit.

For raster data of the moving window analysis, there is a separate file for each spatial scale (radii) as \texttt{Cloud Optimized GeoTIFF} format (\texttt{tif} extension), which contains the value of the Critical Distance for each pixel as signed integer (32 Bit; Rows: 12674; Columns: 27181). 
The rasters contain negative values, which have a special meaning:
\begin{itemize}
  \item values >= 50: Critical Distances
  \item values = -200: Nodata (no settlement area inside the circle area)
  \item values = -300: Water
\end{itemize}
The maximum Critical Distance is the diameter of the respective circle (i.e., twice the radius), as all objects are always assigned to the giant cluster by then at the latest.
Thus, the \texttt{not percolated} status does not occur in the rasters of the moving window analysis.

In addition to the raster files for Critical Distances (\texttt{moving\_win\_r*km\_crit\_dist.tif}), there is also a raster for the proportion of settlement area to terrestrial area for each radius in percent (\texttt{moving\_win\_r*km\_builtup\_share.tif}).
All products are available in the IOER repository.

\begin{table*}[t]
\centering
\begin{tabular}{*{5}{|l}|}
\hline
\multicolumn{5}{|c|}{ \thead{Global Settlement Percolation (GSP) dataset} } \\
\hline
\textbf{Product} & \textbf{File format} & \textbf{File name} & \textbf{Layer name} & \textbf{Spatial scale}\\
\hline
Administrative units & Geopackage & admin\_areas.gpkg & countries & Countries \\ \cline{4-5}
& (Vector data) & & subnations & Sub-nations \\
\hline
Grid cells & Geopackage & grids.gpkg & grid\_0p5\_deg & 0.5$^\circ$ \\ \cline{4-5}
& (Vector data) & & grid\_1\_deg & 1$^\circ$ \\ \cline{4-5}
& & & grid\_2p5\_deg & 2.5$^\circ$ \\ \cline{4-5}
& & & grid\_5\_deg & 5$^\circ$ \\
\hline
Moving window & GeoTIFF & moving\_win\_r1km\_crit\_dist.tif & - & Radius 1 km \\ \cline{3-5}
& (Raster data) & moving\_win\_r2p5km\_crit\_dist.tif & - & Radius 2.5 km \\ \cline{3-5}
& & moving\_win\_r5km\_crit\_dist.tif & - & Radius 5 km \\ \cline{3-5}
& & moving\_win\_r7p5km\_crit\_dist.tif & - & Radius 7.5 km \\ \cline{3-5}
& & moving\_win\_r10km\_crit\_dist.tif & - & Radius 10 km \\ \cline{3-5}
& & moving\_win\_r15km\_crit\_dist.tif & - & Radius 15 km \\
\hline
\end{tabular}
\caption{Taxonomy of the Global Settlement Percolation (GSP) dataset.
Critical Distances are available as three products, i.e.\ administrative units, grid cells and moving window (first, left most column). 
Accordingly, we provide the products either as vector data or as raster data (second column).
The table also provides the file names (third column).
The results of the moving windows analysis are organized into different files.
The results of the other two analyses are in one file each but with multiple layers (fourth column).
In the fifth, right most, column the respective spatial scales are listed.
}
\label{tab:files}
\end{table*}

\begin{figure*}[t]
\centering
\includegraphics[width=\textwidth,keepaspectratio]{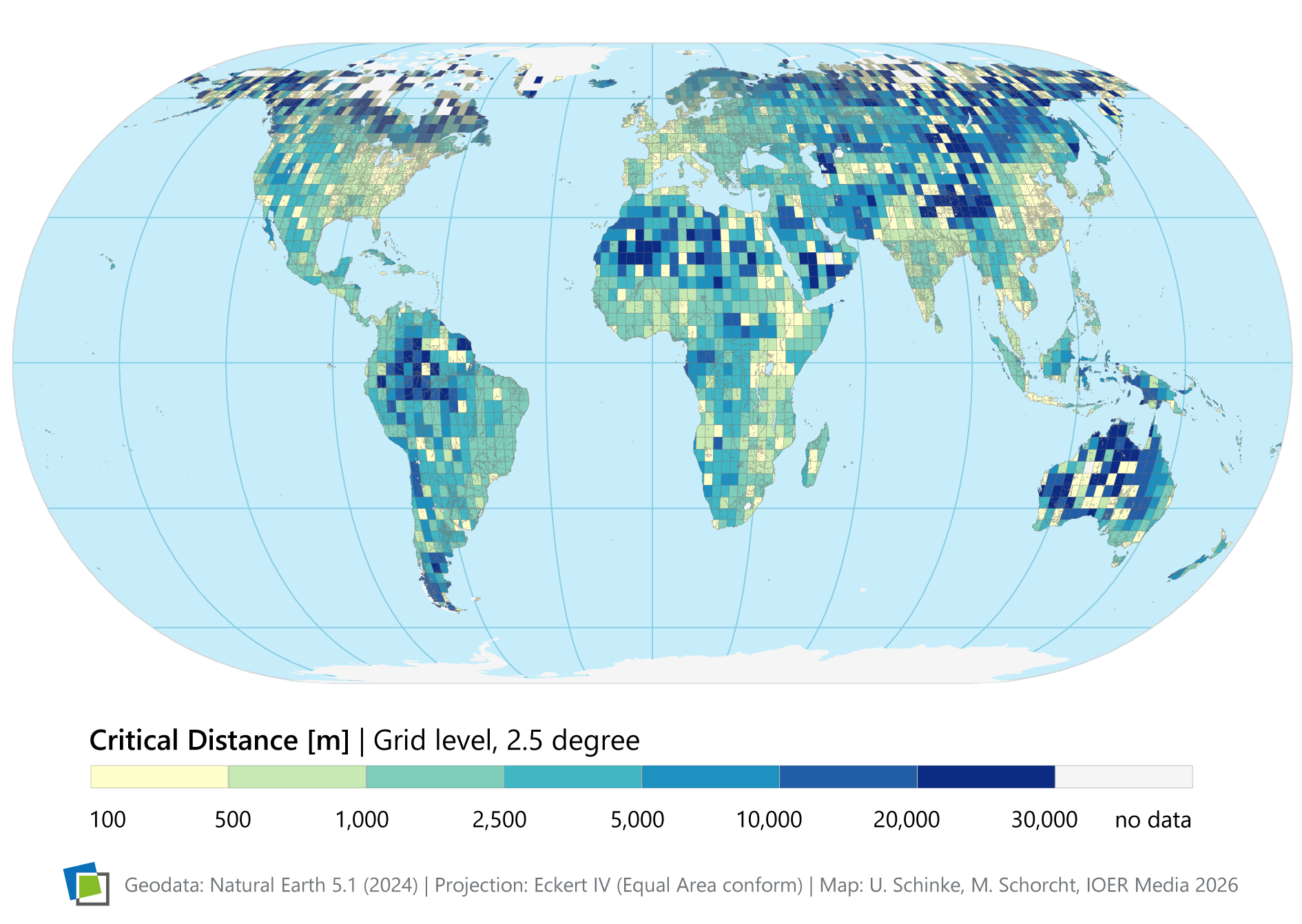}
\caption{Global map of Critical Distances at 2.5$^\circ$ grid.
Every grid cell has been colored according to the Critical Distance that we estimated within it.
Grey: no data.
\label{fig:world_grid2p5_results}}
\end{figure*}

As an example of a global layer, the map of Grid 2.5 degree is shown in Fig.~\ref{fig:world_grid2p5_results}. 
The values of the Critical Distances range from 100\,m to 30,000\,m.
The maximum value of 30,000\,m is in the north of Greenland (Qaasuitsup Kommunia), but other regions such as those of Alaska, Taiga of Russia, Mongolia, West China, and parts of the Amazon basin also have very high Critical Distances. 
In contrast, regions such as the east coast of America, Europe, India, West and East Africa, and East China are characterized by lower Critical Distances.
Global maps of other spatial units can be found in the \hyperref[sec:appendix]{Figures \& Tables} section.

\begin{figure*}[t]
\centering
\includegraphics[width=\textwidth,keepaspectratio]{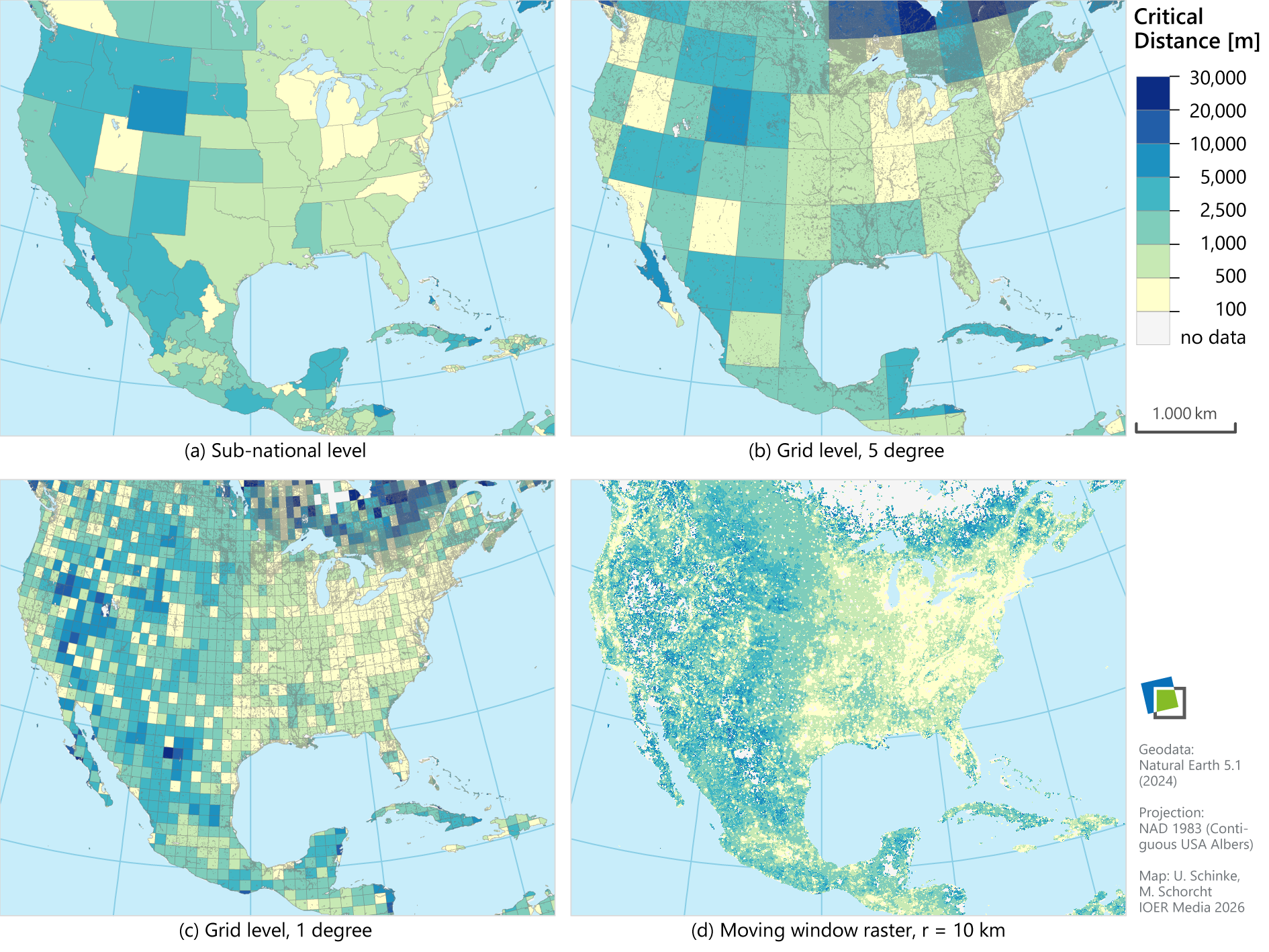}
\caption{Critical Distances in North America.
For various spatial units the values of the Critical Distance are indicated by colors.
The spatial units are (a) sub-national, (b) 5$^\circ$ grid, (c) 1$^\circ$ grid, and (d) moving window with $r=10$\,km.
Grey: no data.
One can see differences in the East and West of the USA.
}
\label{fig:usa_results}
\end{figure*}

Figure~\ref{fig:usa_results} shows a section of North America with different spatial units, i.e.\ sub-national in Fig.~\ref{fig:usa_results}(a), 5$^\circ$ grid in Fig.~\ref{fig:usa_results}(b), 1$^\circ$ grid in Fig.~\ref{fig:usa_results}(c), and 10\,km moving window in Fig.~\ref{fig:usa_results}(d).
It can be seen that relatively low Critical Distances are estimated in densely populated areas such as the entire east and west coast of the United States of America, southern Mexico, or El Salvador. 
In contrast, less densely populated areas such as the Rocky Mountains or the Great Plains have larger Critical Distances.

\section{Correlations with other data}
As no other global calculation of the Critical Distance has yet been carried out,
no comparisons can be made. 
However, in a previous study on the Critical Distance of Germany, which was determined using the Official Building Polygons of Germany (\emph{HU-DE}), a Critical Distance of 830\,m was reported \citep{behnisch_settlement_2019}.
In the present study, however, a Critical Distance of 850\,m is estimated for Germany, which is close to the value already determined. 
This is despite the fact that different input datasets were used (cadastral building coordinates vs.\ objects captured by remote sensing with a resolution of approx.\ 10\,m). 
In addition, in this study the clustering threshold up to 1,000\,m was only calculated in 50\,m increments (\dots{}, 800\,m, 850\,m, 900\,m, \dots{}), whereas 10\,m increments were used in the previous study \citep{behnisch_settlement_2019}.

To get an idea to how complementary the Critical Distances is to other geographical characteristics, we investigated how the Critical Distance relates to other geographical parameters or to what extent these influence the Critical Distance.
Therefore, in addition to correlations with global VIIRS nighttime lights raster layer \citep{VIIRS}, we also examine the correlation with the share of settlement area and the machine-learning based \emph{Human Footprint Index} (\emph{ml-HFI 2019}) from \citet{Keys_hfi_2021}.
We also examine indicators of Urban Sprawl ($WUP_{p}$) and Urban Dispersion ($DIS$) \citep{behnisch_sprawl_2022}.

As these characteristics relate to different spatial scales, we calculate correlations at both small and large spatial scales. 
The nighttime lights and \emph{HFI} are rather local measures and are therefore compared with the results of the moving window raster (resolution of 1\,km with a radius of 1\,km around each pixel).
The \emph{ml-HFI 2019} raster already had a resolution of 1\,km, so it matched the moving window raster well.
The nighttime lights raster, on the other hand, first had to be resampled from approximately 500\,m to 1\,km.
\texttt{nodata} pixels were not included in the analysis.

In contrast, the share of settlement areas, dispersion, and urban sprawl are more related to larger areas and are therefore compared at grid level (0.5~degrees) and at sub-national level.

\begin{figure*}[t]
\centering
\includegraphics[width=\textwidth,keepaspectratio]{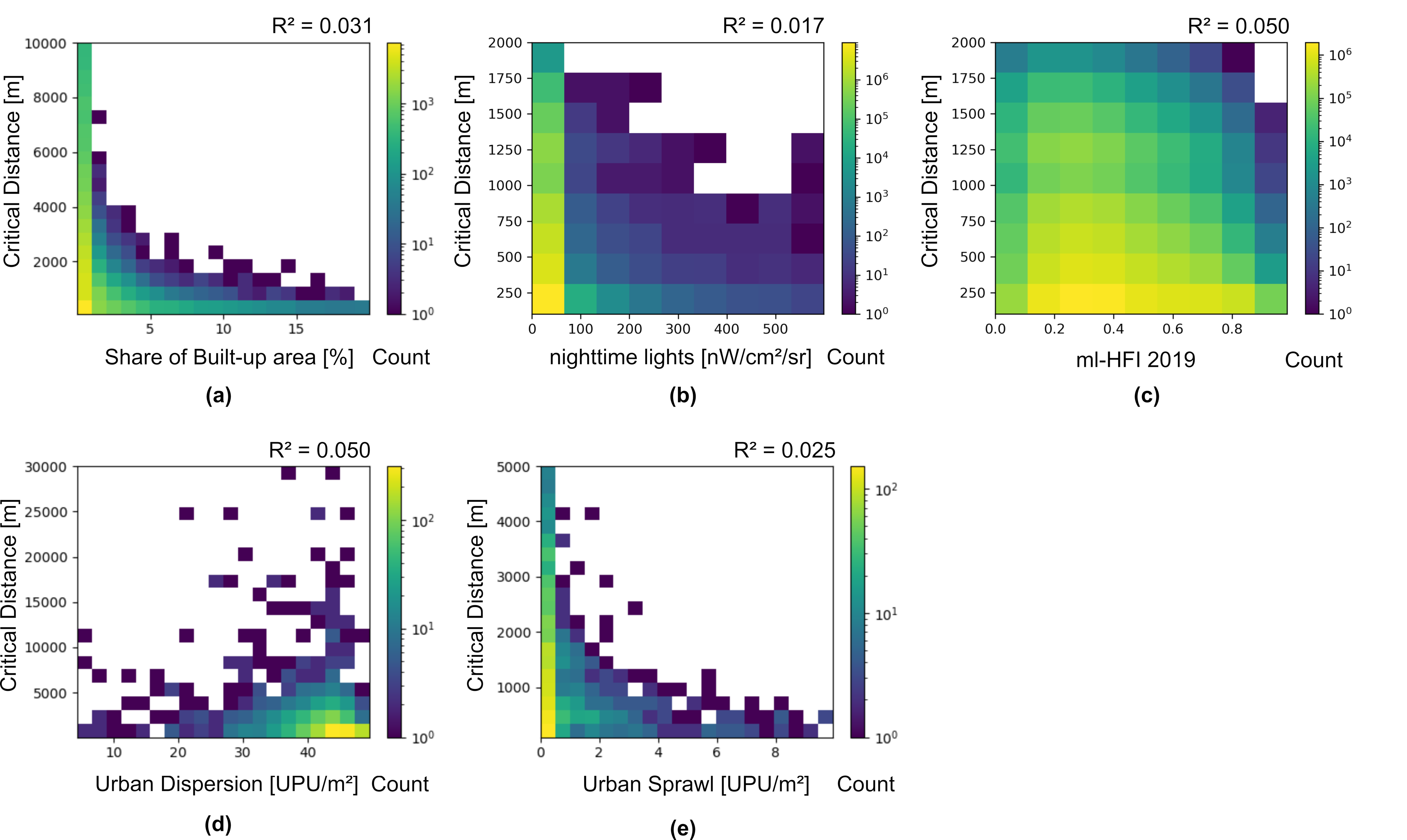}
\caption{Correlation analysis of Critical Distance estimates with various geographical characteristics.
Each panel shows a heat map, i.e.\ a two-dimensional histogram, where the number of counts is visualized by colors (logairthmic scale).
(a) Critical Distance (0.5$^\circ$ grid) vs.\ share of settlement area, 
(b) Critical Distance (moving window, with $r=1$\,km) vs.\ nighttime lights, 
(c) Critical Distance (moving window, with $r=1$\,km) vs.\ Human Footprint Index (HFI), 
(d) Critical Distance (sub-national) vs.\ urban dispersion, and 
(e) Critical Distance (sub-national) vs.\ urban sprawl.
The respective R$^2$-values are shown above each panel.
At best minor correlations are found from which we conclude that the Critical Distance represents a complementary measure to characterize the settlement structure.
}
\label{fig:correlation_plots}
\end{figure*}

Figure~\ref{fig:correlation_plots} consists of cross-plots in the form of 2D histograms, where the number of values falling in a class is shown in color (blue: few; yellow: many; logarithmic color scale).
The values of Fig.~\ref{fig:correlation_plots}(a) do range beyond 20\,\% and the Critical Distance beyond 10\,km, but were disregarded for plotting due to their rarity.
In Fig.~\ref{fig:correlation_plots}(b) and~(c) the value range of the Critical Distance extends from 100\,m to 2,000\,m.
With a radius of 1\,km for the respective circle areas around each pixel of the moving window raster, 2,000\,m is the maximum value, as all settlement pixels within a circle are assigned to a cluster at this value at the latest.
In Fig.~\ref{fig:correlation_plots}(e), the plot was also limited to a Critical Distance of 5,000\,m and urban sprawl of less than 10\,UPU/m$^2$ for the same reason.
Those values with a high Critical Distance all have a low share of settlement area (<1\,\%) or a low UPU/m$^2$ value (<1.25\,\%).
In contrast, the plot in Fig.~\ref{fig:correlation_plots}(d) was not limited, as the values were evenly scattered in all ranges.

From the R$^{2}$ values, an absence of correlations can be seen with the characteristics investigated. 
In the case of built-up area share Fig.~\ref{fig:correlation_plots}(a) the graph shows an inverse relationship.
It can be seen that the Critical Distance decreases as the proportion of built-up area increases, but built-up of less than 2\,\% occur very frequently and co-occur with all Critical Distances. 
Overall, there is no correlation, resulting in an R$^{2}$ value of only 0.031. 
Urban Dispersion Fig.~\ref{fig:correlation_plots}(d) and sprawl Fig.~\ref{fig:correlation_plots}(e) have similar characteristics with an R$^{2}$ value of 0.05 and 0.025.
\emph{HFI} Fig.~\ref{fig:correlation_plots}(c) also has an R$^{2}$ value of 0.050, as Critical Distances of less than 250\,m are evenly distributed across all \emph{HFI} values.

In summary, it can be said that no pronounced correlations were found with regard to the characteristics investigated. 
This underlines that the Critical Distance is an independent measure and thus offers added value with regard to spatial analyses of the connectivity of settlement structures. 
Nevertheless, further studies can investigate which characteristics influence the Critical Distance in order to gain a better understanding of the spatial organization of settlement structures.

\section{Usage Notes}
Both the vector data (\texttt{gpkg} extension) and the raster data (\texttt{tif} extension) are not proprietary file formats and can therefore be used with common GIS programs (e.g.\ \texttt{QGIS}) or programming languages (e.g.\ \texttt{Python} or \texttt{R}).

In case of the moving window analysis, the \texttt{nodata} values (-200) due to the absence of settlements within the given window, can be interpreted as large Critical Distances and filled with large values.
A pragmatic choice is to fill the \texttt{nodata} values with the respective maximum value of the moving window raster, which is the diameter ($2 \times r$).
For example, with a radius of 15,000\,m, the diameter of 30,000\,m can be used.
If users need a complete dataset, filling the \texttt{nodata} values still needs to be done.

The estimated Critical Distances as provided in our Global Settlement Percolation (GSP) data product can be regarded as the \emph{permeability} or \emph{porosity} of the settlement areas.
Since the Critical Distance does not seem to be correlated well with other topological indicators, it can be considered complementary, which makes it interesting for downstream research (see \hyperref[sec:backgroundsummary]{Background \& Summary}).

\conclusions
In summary, we systematically analyze the percolation-like transition of settlements in various spatial units and across the globe.
As a result, we provide the Global Settlement Percolation (GSP) dataset consisting of the respective Critical Distances.
The Critical Distance represents a measure that can be understood as the bottleneck-width between built-up areas.
As such, the Critical Distance is a new and complementary measure that characterizes the morphology and spatial organization of settlements.
Accordingly, we foresee promising downstream use cases for our data, such as addressing landscape structure, accessibility, or strategic land-use planning; enabling research questions such as: How are settlements spatially arranged and connected as elements of the Anthropocene landscape?

Given the role of settlement configuration in shaping ecological connectivity, the GSP dataset also has great potential for ecological applications. While the composition of human settlements is already widely used as a proxy for human presence and associated stressors to wildlife \citep{theobaldGlobalExtentChange2025,tuckerMovingAnthropoceneGlobal2018b}, global-scale analyses on how the spatial arrangement of human-modified versus open areas facilitate or impede ecological processes such as the movements of terrestrial wildlife have been hampered by a lack of suitable data characterising configuration. While hundreds of landscape configuration metrics (e.g., aggregation index, nearest-neighbor distance, fractal dimension \citep{cushmanParsimonyLandscapeMetrics2008} have been developed, high redundancy and collinearity among these metrics make it difficult to identify a consistent, transferable approach for large-scale, multi-species, and/or cross-regional analyses. As a result, findings from local and regional studies are rarely comparable or generalizable, which is a major hindrance for developing insights into the impacts of settlement configuration on ecological functioning \citep{cushmanParsimonyLandscapeMetrics2008,calabreseComparisonshoppersGuideConnectivity2004}. The Critical Distance metric addresses this gap by providing a scalable, numerically continuous, and integrative index of landscape configuration. 
The spatial configuration of settlements influences the ability of organisms, ecological processes, or habitats to remain connected or not.

\codedataavailability{
The dataset is available at the repository ioerDATA.
During the review process, the dataset is available at a publicly accessible preview URL.

There is an easy-to-use tool as a \texttt{Docker\,Compose} environment (or native as \texttt{Python} and \texttt{PostgreSQL} scripts) on \texttt{Github} for calculating the Critical Distances.
The code, including detailed documentation, can be found at:
\textcolor{blue}{\url{https://github.com/schorcht-ioer/critical_mapper}\, .}

In this \texttt{Docker} environment, a \texttt{PostgreSQL} database is automatically set up, along with the developed {Python} scripts and other requirements. 
It is also permits to configure the number of usable CPU cores and RAM utilization so that high-performance parallel processing is possible. 
In addition to calculating the Critical Distance, this tool can also be used solely for clustering large raster and vector datasets.
}

\appendix

\appendixfigures

\appendixtables

\begin{figure*}[t]
\centering
\includegraphics[width=0.8\textwidth,height=\textheight,keepaspectratio]{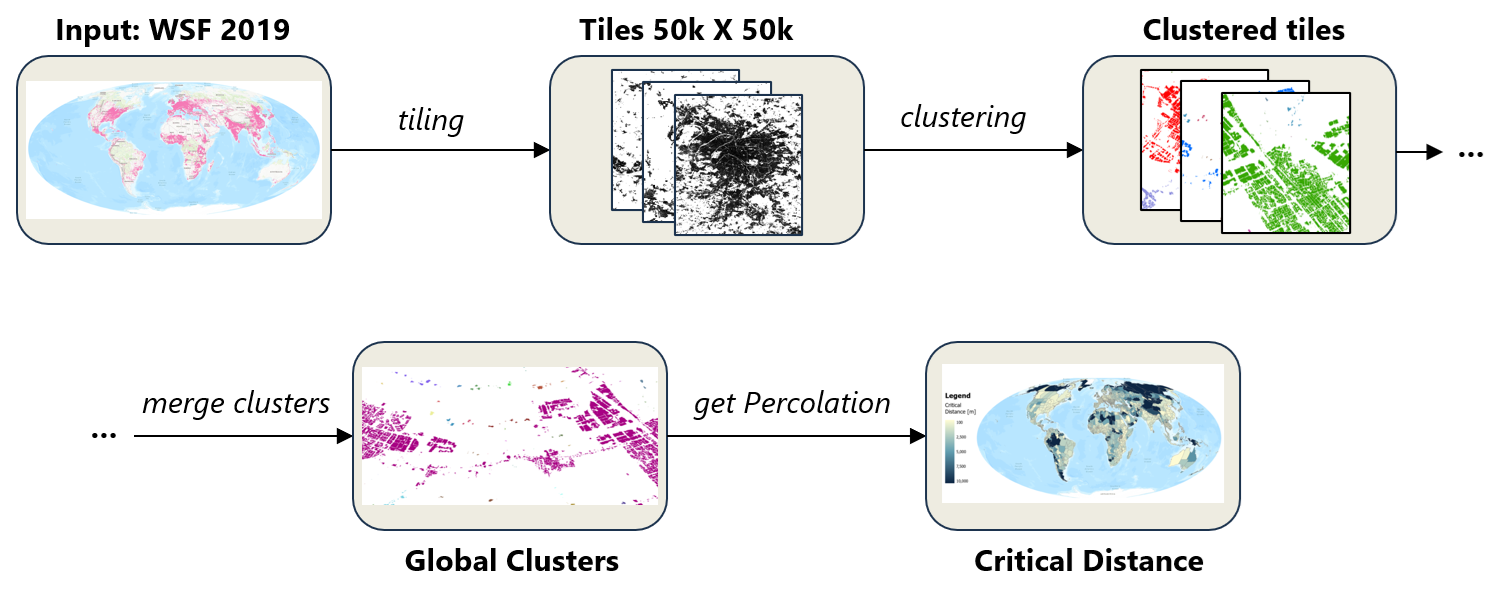}
\caption{Workflow of data processing and steps until determination of Critical Distances.
}
\label{fig:workflow}
\end{figure*}

\begin{table*}[ht]
\centering
  \begin{tabular}{|l|l|c|c|}
    \hline
    \thead{Attribute name} & \thead{Description} & \thead{Unit} & \thead{Type} \\
    \hline
    gid & Unique identifier & - & integer \\
    \hline
    name & Name of administrative area $^{1}$ & - & text\\
    \hline
    code & Code of administrative area $^{2}$ & - & text\\
    \hline
    critical\_distance & The Critical Distance of the spatial unit & m & integer \\
    \hline
    builtup\_area\_sqkm & Settlement area (geodetic) of the administrative area & km$^{2}$ & double \\
    \hline
    builtup\_share & Share of Settlement area of the spatial unit & \% & double \\
    \hline
    terrestrial\_area\_sqkm & Terrestrial area (geodetic) of the administrative area & km$^{2}$ & double \\
    \hline
    status & Status of the Critical Distance calculation & - & text \\
    \hline
  \end{tabular}
\caption{\label{tab:attributes} Attributes of the vector data (Administrative units and Grid cells). $^{1}$ Source attribute of \emph{Natural Earth} dataset: Admin 0 (countries) → ADMIN; Admin 1 (subnations) → name;  
Grid cells have the following name patterns: [grid size in degree][longitude][latitude] (from the bottom left corner); $^{2}$ Is not available with the grid layers;
}
\end{table*}

\begin{figure}[t]
\centering
\includegraphics[width=\columnwidth]{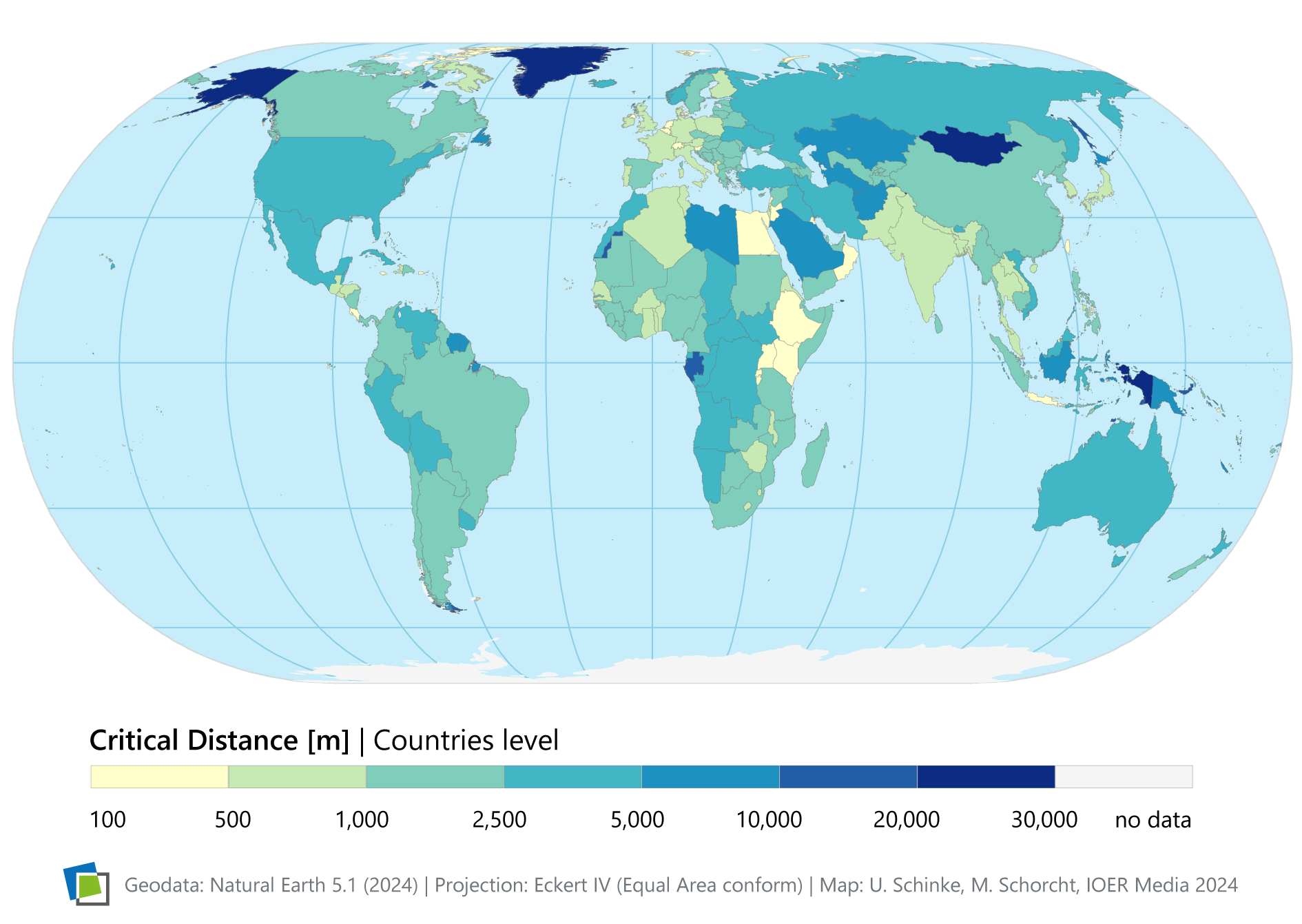}
\caption{
Global map of Critical Distances at national scale.
Every country has been colored according to the Critical Distance that we estimated within it. 
Grey: no data.}
\label{fig:world_countries_results}
\end{figure}

\begin{figure}[t]
\centering
\includegraphics[width=\columnwidth]{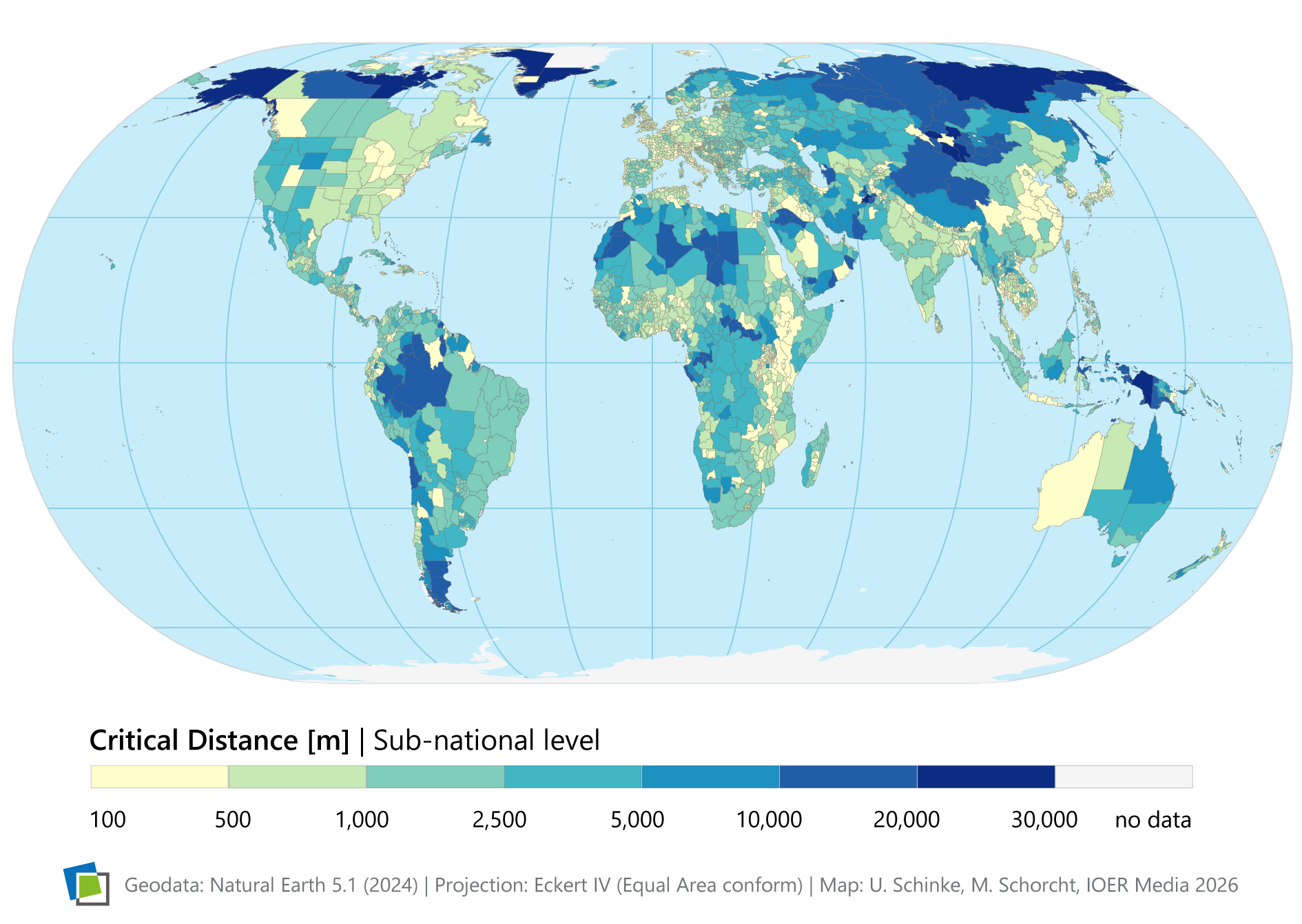}
\caption{Global map of Critical Distances at sub-national scale.
Every sub-national unit has been colored according to the Critical Distance that we estimated within it.
Grey: no data. }
\label{fig:world_subnations}
\end{figure}

\begin{figure}[t]
\centering
\includegraphics[width=\columnwidth]{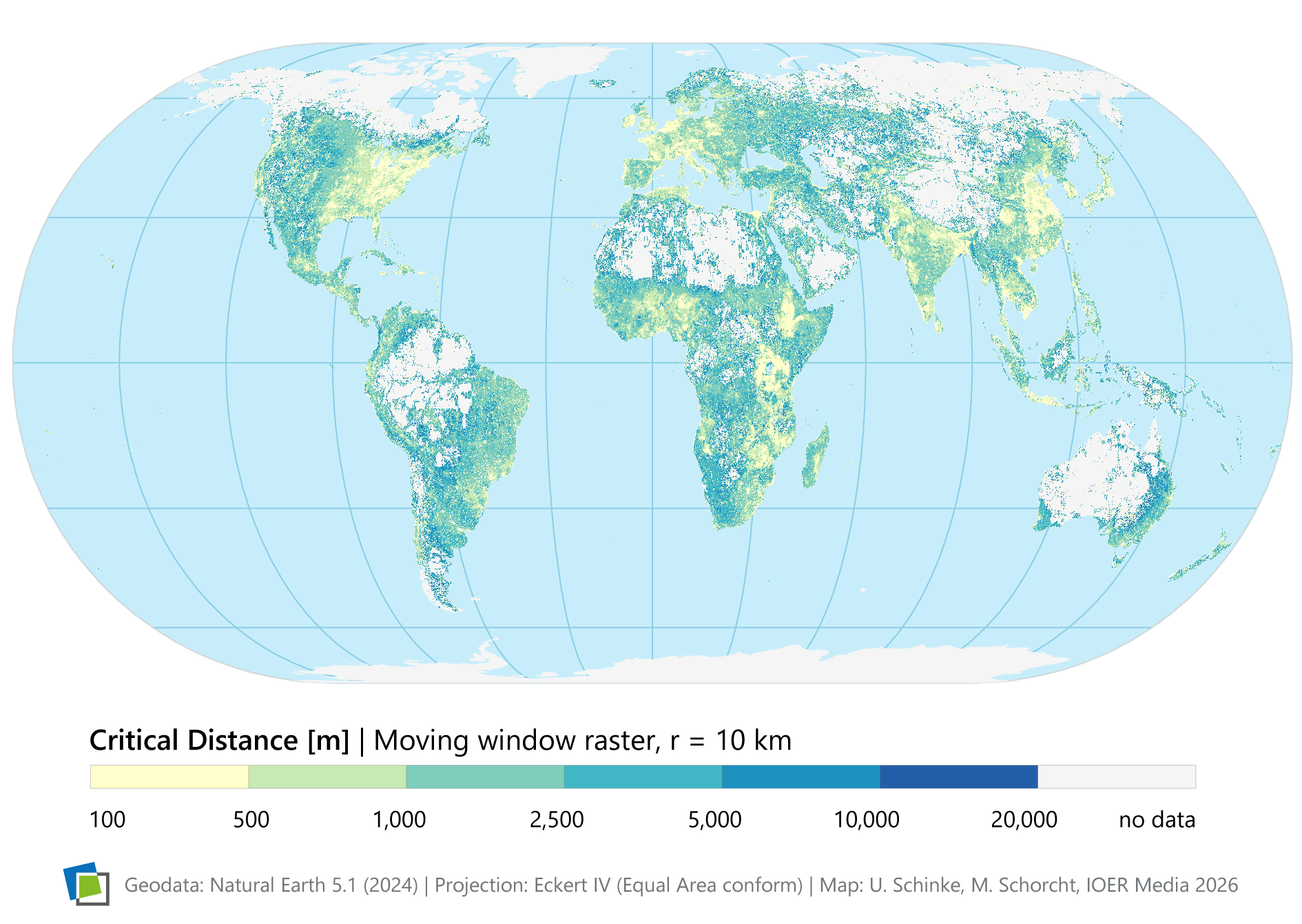}
\caption{Global map of Critical Distances within moving windows of radius $r=10$\,km.
Every cell has been colored according to the Critical Distance that we estimated within it.
Grey: no data. }
\label{fig:world_movingwindow_results}
\end{figure}

\noappendix 

\authorcontribution{
Conceptualization: M.S., M.B. L.T.B., A.-K.B., T.M., D.R.;
Methodology: M.S., D.R.;
Software: M.S.;
Validation: M.S.;
Formal analysis: M.S.;
Investigation: all;
Resources: M.S.;
Data Curation: M.S.;
Writing - Original Draft: M.S., D.R.;
Writing - Review \& Editing: all;
Visualization: M.S., R.L.F, W.X.;
Supervision: M.B., T.M., D.R.;
Project administration: M.B., T.M., A.-K.B.;
Funding acquisition: M.B., L.T.B., T.M., D.R.\\
}

\competinginterests{The authors declare no competing interests.}

\begin{acknowledgements}
We thank Leibniz Association for funding the project ``Landscape Criticality in the Anthropocene -- Biodiversity, Renewables and Settlements'' (CriticaL) within which this research was conducted.
We would also like to thank J.\ G\"ossel for hardware and software support at IOER.
Additionally, D.R.\ would like to thank the German Research Foundation (DFG) for funding the projects UPon (\#451083179) and Gropius (\#511568027).
\end{acknowledgements}

\end{document}